\documentclass[prd,twocolumn,aps,amsmath,nofootinbib,superscriptaddress]
{revtex4}

\usepackage{graphicx}
\usepackage{hyperref}
\usepackage{amssymb}
\usepackage{bm}

\newcommand{\bea}{\begin{eqnarray}}
\newcommand{\eea}{\end{eqnarray}}
\newcommand{\beq}{\begin{equation}}
\newcommand{\eeq}{\end{equation}}
\newcommand{\nn}{\nonumber}
\newcommand{\comment}[1]{}

\def\Hp{H_{\rm p}}
\def\Hd{H_{\rm d}}

\begin{document}


\preprint{} 

\title{A signature of anisotropic bubble collisions}

\author{Michael P.~Salem}
\affiliation{Institute of Cosmology, Department of Physics and Astronomy,\\
Tufts University, Medford, MA 02155}


\begin{abstract}
Our universe may have formed via bubble nucleation in an 
eternally-inflating background.  Furthermore, the background may have 
a compact dimension---the modulus of which tunnels out of a metastable 
minimum during bubble nucleation---which subsequently grows to become 
one of our three large spatial dimensions.  When in this scenario our 
bubble universe collides with other ones like it, the collision 
geometry is constrained by the reduced symmetry of the tunneling 
instanton.  While the regions affected by such bubble collisions still 
appear (to leading order) as disks in an observer's sky, the centers of 
these disks all lie on a single great circle, providing a distinct 
signature of anisotropic bubble nucleation.
\end{abstract}


\maketitle

\section{Introduction}
\label{sec:introduction}

String theory argues for the existence of an enormous landscape of 
metastable, positive-energy vacua (in addition to other states) 
\cite{BP,S03}.  If any such vacuum is obtained in spacetime, it 
expands exponentially and without bound:  while one expects quantum 
transitions, in the form of bubble nucleation, from any one such 
vacuum to another \cite{CDL}, in metastable states the resulting 
bubbles do not percolate \cite{Guth81}.  Thus emerges a view of 
cosmology that sees our universe as part of the inside of a bubble, 
which nucleated within some eternally-inflating background, in 
which other bubbles endlessly nucleate and collide \cite{GV97}. 

The string landscape corresponds in part to the various ways to 
compactify the higher-dimensional fundamental theory down to a 
(3+1)-dimensional spacetime such as we observe.  Yet there should 
also exist metastable compactifications with fewer and more large 
spatial dimensions, in which case we expect these vacua to also 
play a part in the above cosmology.  The consequences of this have 
only just begun to be explored \cite{B-PS-PV,CJR,B-PS-PV-2,B-PS,
GHR,ACN,B-PS2}.  We here focus on the scenario where the vacuum 
in which our bubble nucleates has an additional compactified 
dimension---the modulus of which tunnels out of a metastable 
minimum during bubble nucleation---which subsequently grows to 
become one of our three large spatial dimensions.  Then the 
reduced symmetries of the parent vacuum (relative to the local, 
SO(3,1)-symmetric asymptotic geometry in the bubble) imply 
anisotropic initial conditions for the evolution of our universe.

Some signatures of this scenario have been explored in
\cite{B-PS,GHR,B-PS2}, where ultimately it is found that an 
induced quadrupole of order the present curvature parameter, 
$\Omega_{\rm c}^0$, places severe observational constraints on 
such effects as statistically-anisotropic perturbations in the 
cosmic microwave background (CMB), and spatial variations in 
the observed luminosity of standard candles, both of which are 
suppressed relative to leading-order terms by $\Omega_{\rm c}^0$.
Yet these are not the only possible observational signatures
of anisotropic bubble nucleation.

We here study a certain class of bubble collisions, namely 
those involving two bubbles of the type that contains our 
universe, assuming such bubbles nucleate as described above, 
via metastable modulus decay.  With such bubbles the tunneling
instanton is independent of the coordinate of the compact 
dimension, so that when two such bubbles collide, the locus of
events at the ``point'' of first contact spans the entire 
compact dimension.  This, along with the reduced symmetry of 
the parent vacuum, constrains the geometry of such bubble 
collisions.  Setting aside order $\Omega_{\rm c}^0$ affects due
to background anisotropy, the regions affected by these 
collisions project onto disks on the CMB sky.  However, unlike 
the case with a (3+1)-dimensional parent vacuum, the centers of 
these disks all lie on a single great circle.  In the 
circumstance where at least three such collisions are observed,
this provides a distinct signature of anisotropic bubble 
nucleation.

Detection of effects of bubble collisions requires that the 
number of $e$-folds $N_e$ of inflation in our bubble not be too 
large.  In the model of anisotropic bubble nucleation that we 
consider here, observational constraints on $N_e$ are tighter 
than in the isotropic scenario, since absent any fortuitous 
cancellations coming from other (still unexplored) large-scale 
effects, the induced quadrupole mentioned above requires 
$\Omega_{\rm c}^0\lesssim 10^{-5}$.  However, because
the initial effects of bubble collisions can be much larger 
than the spatial curvature on scales now entering the horizon,
there is hope to detect signals of bubble collisions even when
the present-day curvature parameter is well below the level of 
cosmic variance \cite{worldscollide,Johnson}.

The study of (3+1)-dimensional bubble collisions in a 
(3+1)-dimensional parent vacuum has a long history, for an 
excellent summary see \cite{AJreview} (and references therein).  
Most aspects of the present analysis follow directly from 
ideas and techniques developed in that scenario; this work 
in particular benefits from \cite{GGV,AJS,FKNS,Dahlen,AJreview}.

The remainder of this paper is organized as follows.  In 
Section \ref{sec:background} we briefly review a toy model of 
modulus stabilization, providing for an effectively 
(2+1)-dimensional parent vacuum, and describe the tunneling 
instanton that connects to our asymptotically 
(3+1)-dimensional daughter bubble.  The  basic features of a 
collision between two such daughter bubbles in such a parent 
vacuum is described in Section \ref{sec:collision}, where we 
work out the collision-affected region in an observer's CMB sky.  
In Section \ref{sec:dist} we discuss the distribution of 
positions and angular scales of such regions, under some 
plausible guesses about the setup.  Concluding remarks are 
provided in Section \ref{sec:conclusions}.

\section{Modulus (meta)stabilization and tunneling instanton}
\label{sec:background}

A toy model to implement anisotropic bubble nucleation is 
described in \cite{B-PS}.  We here briefly review the model, and
discuss the tunneling instanton, to motivate essential features 
of the geometry used in Section \ref{sec:collision}.  The model
uses the winding number of a complex scalar field to stabilize 
the size of the compact dimension; in particular the 
(3+1)-dimensional action is\footnote{String theory indicates 
that the (3+1)-dimensional vacuum itself has six or seven 
compact dimensions, however throughout this paper we consider
the associated moduli fields to be non-dynamical spectators in
all of the processes of interest.}
\beq
S_{\rm 4d} = \int\! \sqrt{-g}\,d^4x 
\left[ \frac{1}{16\pi G}\left( R - 2\Lambda\right) +
{\cal L}_\varphi\right] ,
\label{4daction}
\eeq  
where $g$ is the determinant of the (3+1)-dimensional metric 
$g_{\mu\nu}$, $R$ is the corresponding Ricci scalar, $\Lambda$ 
is a cosmological constant, and 
\beq
{\cal L}_\varphi=-\frac{1}{2}K(\partial_\mu\varphi^*\partial^\mu\varphi) 
- \frac{\lambda}{4}\left(|\varphi |^2-m^2\right)^2 \,,
\eeq  
where $\varphi$ is the scalar, for which we allow a non-canonical 
``kinetic'' function specified by $K$.  The other terms are 
constants.  Other degrees of freedom, for instance the inflaton 
and the matter fields of the Standard Model, are assumed to be 
unimportant during the tunneling process, and are absorbed into 
$\Lambda$ (and/or $g$ and $R$).

The stabilization of the volume modulus is studied by starting 
with a metric ansatz with line element
\beq
ds^2 = e^{-\Psi}\,\overline{g}_{ab}\,dx^adx^b 
+ L^2e^{\Psi}\,dz^2 \,,
\label{ansatz}
\eeq   
where $\Psi$ represents the modulus field.  The effective 
(2+1)-dimensional metric $\overline{g}_{ab}$ and the modulus 
$\Psi$ are both taken to be independent of $z$.  Meanwhile, the 
compact dimension $z$ is defined using periodic boundary 
conditions, with $-\pi<z<\pi$, so that it has the topology of a 
circle with physical circumference $2\pi L\,e^{\Psi/2}$.  Note 
that we have introduced the following notation.  Any quantity 
defined explicitly within the effective (2+1)-dimensional 
theory (the theory with the $z$ dimension integrated out), 
such as the (2+1)-dimensional metric, is marked with an 
overline.  Whereas Greek indices are understood to run over all 
dimensions, Latin indices are understood to run over all but 
the $z$ dimension.

The equations of motion permit solutions of the form
\beq
\varphi\approx\left(m^2 -\frac{n^2}{\lambda L^2}K'
e^{-\Psi}\right)^{\!1/2}\! e^{inz} \,,
\label{phisol}
\eeq
if we take 
\beq
m^2 \gg \frac{n^2}{\lambda L^2}K'e^{-\Psi} \,. 
\label{etaapprox}
\eeq
Here $n$ is an integer and $K'\equiv dK(X)/dX$, with 
$X\equiv\partial_\mu\varphi^*\partial^\mu\varphi=(n^2m^2/L^2)\,
e^{-\Psi}$.  After integrating by parts, we find the effective 
(2+1)-dimensional action 
\beq
S_{\rm 3d} = \int\! \sqrt{-\overline{g}}\,d^3x\left[ 
\frac{1}{16\pi \overline{G}}\,\overline{R}
-\frac{1}{2}\partial_a\psi\partial^a\psi 
-\overline{V}(\psi)\right] ,
\label{3daction}
\eeq
where $\overline{G}\equiv G/(2\pi L)$, $\psi\equiv \Psi/\alpha$, 
with $\alpha=\sqrt{16\pi \overline{G}}$, 
and 
\beq
\overline{V}(\psi) = \frac{\Lambda}{8\pi \overline{G}}\,
e^{-\alpha\psi}+\frac{1}{2}e^{-\alpha\psi} \overline{K}(X(\psi)) \,.
\eeq

\begin{figure}[t!]
\includegraphics[width=0.4\textwidth]{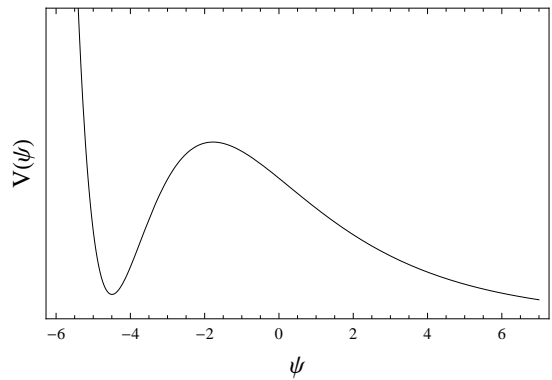}\phantom{SS} 
\caption{\label{fig:V}The effective potential of the modulus 
field $\psi$.}
\end{figure}

To stabilize the modulus it is necessary to introduce at least 
three terms into (a polynomial) $\overline{K}(X)$; since we are here
simply interested in producing a viable model, we use the ad hoc
model 
\beq
\overline{K}(X) = 2 \pi L\left( X + \kappa_2 X^2 
+ \kappa_3 X^3\right)\,,
\label{Kdef}
\eeq  
where $\kappa_2$ and $\kappa_3$ are constants.  The resulting
effective potential $\overline{V}(\psi)$ is displayed in 
Figure~\ref{fig:V} (using the same parameter values as in 
\cite{B-PS}).  There is a metastable minimum at some value 
$\psi=\psi_{\rm p}$, corresponding to the parent vacuum state, 
which appears as (2+1)-dimensional de Sitter space on scales 
much larger than $2\pi L\,e^{\alpha\psi_{\rm p}/2}$.  The 
daughter vacuum is created when $\psi$ tunnels through the 
barrier, to some value $\psi=\psi_{\rm d}$, after which $\psi$ 
accelerates from rest and rolls down the potential, with 
$\psi\to\infty$ as time $x^0\to\infty$.  For more details see 
\cite{B-PS}.

The semi-classical theory of vacuum decay via bubble nucleation in
(3+1) dimensions is laid out in \cite{CDL}; bubble nucleation in 
our scenario proceeds analogously.  The essential difference is the
reduced symmetries in the parent vacuum.  In our scenario the 
tunneling field is also the modulus, which is independent of the 
compact $z$ dimension (by hypothesis).  Thus the instanton must be 
independent of $z$, and the compact dimension acts as a bystander 
to the tunneling process.  The remaining geometry is SO(2,1) 
invariant, allowing one to follow the methods of \cite{CDL}, which 
starts with an SO(3,1) invariant geometry, by merely changing some 
numerical factors.

Let us briefly sketch the procedure to help describe the results.  
The tunneling instanton is found by studying the O(3)-symmetric
Euclidean line element (ansatz) of the (2+1)-dimensional parent 
vacuum,
\beq
ds^2 = d\chi^2+\rho^2(\chi)
\left[d\xi^2+\cos^2(\xi)\,d\phi^2 \right] \,.
\label{Emetric}
\eeq
The scale factor $\rho$ and the tunneling field $\psi$ obey the 
``inverted-potential'' Euclidean equations of motion, 
\bea
\frac{\dot{\rho}^2}{\rho^2} -\frac{1}{\rho^2} 
\!&=&\! 8\pi \overline{G} \left(\frac{1}{2}\dot{\psi}^2 
- \overline{V}\right) \label{E}\\
\ddot{\psi} + 2\frac{\dot{\rho}}{\rho}\dot{\psi} 
\!&=&\! \overline{V}' \,,\phantom{\bigg(\bigg)} \label{F}
\eea
where the dot and prime denote differentiation with respect to 
$\chi$ and $\psi$ respectively.  In the tunneling solution, as
$\chi$ runs from zero to some maximum value $\chi_{\rm max}$, 
$\psi$ interpolates from $\psi_{\rm d}$ to some value 
$\tilde{\psi}_{\rm p}$ not far from the parent vacuum state 
$\psi_{\rm p}$, with $\dot{\psi}\to 0$ in both limits.  In the 
limit $\chi\to\chi_{\rm max}$, the scale factor approaches 
\beq
\rho(\chi)\to \Hp^{-1}\sin(\Hp\chi+\delta_{\rm p}) \,,
\eeq   
where 
$\Hp^2\equiv 8\pi\overline{G}\,\overline{V}(\tilde{\psi}_{\rm p})$ 
and the phase $\delta_{\rm p}$ is determined by boundary conditions.  
An analogous solution applies in the limit $\chi\to 0$.

The Lorentzian solution is determined by analytic continuation,
$\xi\to i\xi$, which gives the line element 
\beq
ds^2 = d\chi^2+\rho^2(\chi)
\left[-d\xi^2+\cosh^2(\xi)\,d\phi^2 \right] \,.
\label{Lmetric}
\eeq
In the appropriate limits of $\chi$ the above solution corresponds 
to a slicing of de Sitter space.  The instanton boundary $\chi=0$ 
is now the future lightcone of the origin of coordinates, and 
serves as the boundary of the nucleated bubble geometry.  The
initial conditions for the bubble are determined by matching, 
and can be obtained by analytic continuation of (\ref{Lmetric}).  
Taking $\chi\to i\chi$ and $\xi\to \xi-i\pi/2$ gives the line
element
\bea
ds^2 \!\!&=&\!\! -\,e^{-\Psi(\chi)}d\chi^2 + e^{-\Psi(\chi)}
\rho^2(\chi)\left[d\xi^2+\sinh^2(\xi)\,d\phi^2\right] \nn\\
& &\!\! +\, L^2e^{\Psi(\chi)}\, dz^2 \,,
\label{bubblegeom0}
\eea
where we have restored the conformal factor 
$e^{-\Psi}=e^{-\alpha\psi}$ of the original metric ansatz 
(\ref{ansatz}), and have revealed the compact dimension $z$.
Note that while the analytically-continued instanton solutions 
for $\Psi=\psi/\alpha$ and $\rho$ describe the bubble geometry 
near the instanton boundary, the future evolution of the 
metric components of (\ref{bubblegeom0}) is determined by the
matter content of the bubble.

\section{Anisotropic bubble collisions}
\label{sec:collision}

We find it convenient to parameterize the geometry of the parent
vacuum using a flat de Sitter slicing on the inflating 
submanifold, with line element of the form
\beq
ds^2= A^2(\sigma)\!
\left(-d\sigma^2+d\rho^2+\rho^2d\phi^2\right)+B^2dz^2 ,
\label{flatdS}
\eeq
where $B$ is a constant, setting the size of the compact $z$ 
dimension ($-\pi<z<\pi$), and $A(\sigma)=-1/\Hp\sigma$, with 
$\Hp$ being the de Sitter Hubble rate of the parent vacuum (we 
everywhere ignore the back-reaction of bubble walls on the 
metric).  Meanwhile, the interior of a given bubble can be 
described with a line element of the form
\bea
ds^2=a^2(\eta)\!\left[-d\eta^2+d\xi^2+\sinh^2(\xi)\,d\phi^2\right]\!
+b^2(\eta)\,dz^2 , \,\,\, \label{openFRW}
\eea
with $\eta$ foliating hypersurfaces of constant density.  The 
scale factors $a$ and $b$ are determined by the matter content 
of the bubble, with the instanton boundary conditions setting 
$a\to 0$ and $b\to B$ as $\eta\to -\infty$.  

We assume the energy density in the bubble is initially 
dominated by the inflaton potential, and that it behaves 
essentially as cosmological constant, with asymptotic Hubble
rate $\Hd$.  The solution for the scale factors, up until the
end of inflation, is then
\bea
a(\eta) &=& \Hd^{-1}{\rm csch}
\Big[\ln(\Hp/\Hd)-\eta\Big] \label{sfa} \\
b(\eta) &=& B \coth\!\Big[\ln(\Hp/\Hd)-\eta\Big] \,,
\label{sfb}
\eea
where the offset $\ln(\Hp/\Hd)$ is introduced to facilitate 
matching at the instanton boundary.  In particular, at very early 
times, $-\eta\gg\ln(\Hp/\Hd)$, this solution becomes 
$a(\eta)\approx (2/\Hd)\,e^\eta$, which corresponds to the 
early-time limit of $a(\eta)=\Hp^{-1}{\rm csch}(-\eta)$, the 
scale factor solution for open de Sitter spacetime with 
asymptotic Hubble rate $\Hp$.  This means that we can extend the 
flat de Sitter chart of the parent vacuum (slightly) into the 
bubble, and match coordinates using the (early-time limit of the) 
standard flat--to--open de Sitter coordinate transformations.  
For matching at the future lightcone of a bubble nucleating at
$(\sigma,\,\rho)=(-\Hp^{-1},\,0)$, this gives
\bea
-\!\Hp\sigma \!&=&\! \left(1+e^{\,\xi+\eta}\right)^{-1} 
\label{trans1} \\
\Hp\rho \!&=&\! \left(1+e^{-\xi-\eta}\right)^{-1}\,, 
\label{trans2}
\eea
where we have also used $\xi+\eta =$ constant, and therefore 
$\xi\to\infty$ as $\eta\to-\infty$, as will be appropriate when
we apply these transformations later.  The coordinates $\phi$ 
and $z$ are matched trivially at the boundary.

Note that the evolution of the bubble begins with a period of 
curvature domination, with the subsequent period of inflation 
beginning at $\eta-\ln(\Hp/\Hd)\approx -1$.  During inflation 
$\dot{a}/a$ and $\dot{b}/b$ rapidly converge to generate a 
locally SO(3,1)-symmetric geometry, after which the bubble 
interior may follow the standard big bang cosmology.  (While it 
is possible for effects of the background anisotropy to become 
significant at late times \cite{DD}, observational constraints 
limit such effects to be very small, and we ignore them here.)  
Empirically, the number of $e$-folds of inflation is large, 
meaning inflation ends at time $\eta_{\rm RH}\approx\ln(\Hp/\Hd)$.  
The subsequent evolution of $\eta$ is dominated by its growth 
between recombination and the present, during which $\eta$ 
changes by \cite{B-PS}
\beq
\Delta\eta\equiv r_\star\approx 6.1\sqrt{\Omega_{\rm c}^0} \,,
\label{rstar}
\eeq
where $\Omega_{\rm c}^0$ is the present-day curvature parameter.
This is observationally constrained to be small,
$\Omega_{\rm c}^0\lesssim 6.3\times 10^{-3}$ \cite{WMAP}, and
in the context of anisotropic bubble nucleation, absent 
fortuitous cancellations from other large-scale effects, an 
induced quadrupole constrains it to be even smaller,
$\Omega_{\rm c}^0\lesssim 10^{-5}$ \cite{DD}.  All this is to 
obtain the time of a present observer,
\beq
\eta_0 \approx \ln(\Hp/\Hd) + r_\star \,.
\eeq

\begin{figure}[t!]
\includegraphics[width=0.4\textwidth]{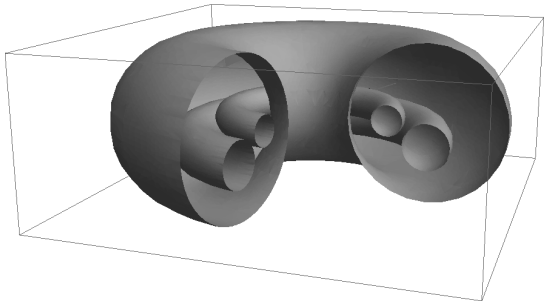}\phantom{SSS} 
\caption{\label{fig:cartoon}Representation of a spatial slice of 
the parent vacuum, involving two bubbles nearing collision (see main
text).}
\end{figure}

Before proceeding to calculations, let us discuss the 
qualitative features of the bubble collision geometry.  
Figure \ref{fig:cartoon} represents two anisotropic bubbles 
nearing collision.  The largest torus represents a spatial 
section of the parent vacuum:  the coordinate that wraps 
around the vertical axis is the compact $z$ dimension, while 
the radial/vertical cross sections are subsets of the 
($\rho$, $\phi$) plane (one should imagine all tori as flat, 
$S^1\!\times\! S^1$).  The two smaller tori represent bubbles 
of daughter vacuum, with the coordinates $\phi$ and $z$ 
aligned with those of the parent vacuum at the boundary (note 
that the radial/vertical slices are not spatial slices of the 
open-FRW coordinates inside the bubbles).  Distances in the 
figure reflect comoving coordinate separations:  on subsequent 
spatial slices the metric on the radial/vertical cross 
sections of the parent vacuum expands according to the scale 
factor $A$, while the azimuthal metric component of the 
larger torus is a constant, $B$.  The azimuthal metric inside 
the bubbles expands with the scale factor $b$, and radial/vertical
cross sections inside the bubbles expand with $a$.  

The bubbles expand into the parent vacuum at a rate approaching the 
speed of light; however the intervening parent vacuum also expands.  
In comoving coordinates, this translates to bubble radii that 
asymptotically approach bubble-nucleation-time-dependent constants.  
Two bubbles collide if they nucleate sufficiently close to each 
other, or at sufficiently early times, for their asymptotic cross 
sections to overlap.  

A crucial feature of Figure \ref{fig:cartoon} is that the bubble
walls are independent of the compact dimension, i.e. the bubbles 
reflect the toroidal symmetry of the diagram.  This is a property of 
the tunneling instanton discussed in Section \ref{sec:background}, 
and it implies that the ``point'' of first contact between the two 
bubbles is in fact a ring, spanning the length of the compact 
dimension.  Our primary interest is to understand the effect of
this.  

\begin{figure}[t!]
\includegraphics[width=0.4\textwidth]{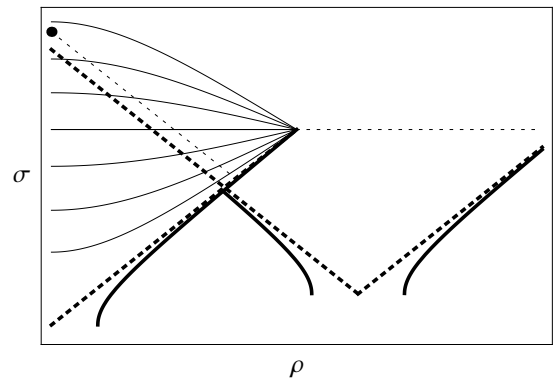}\phantom{SSS} 
\caption{\label{fig:collide}Representation of a spacetime slice of
a bubble collision, suppressing the ($\phi$, $z$) torus (see main 
text).}
\end{figure}

A more traditional representation of a bubble collision is given
in Figure \ref{fig:collide}.  Here we suppress the ($\phi$, $z$) 
torus at every point, displaying the ($\sigma$, $\rho$) plane of 
the parent vacuum and ($\eta$, $\xi$) planes of the bubbles.  (The 
radial/vertical cross sections of Figure \ref{fig:cartoon} 
correspond to horizontal lines in this figure.)  Here the solid dot 
specifies the position of the observer, the light dotted line the 
observer's past lightcone, the dark dotted lines the future 
lightcones of the bubble nucleation events (technically, the 
tunneling instanton boundaries), the dark solid lines bubble domain 
walls, and the solid curves in the observer's bubble surfaces 
of equal bubble time $\eta$.    

The bubble collision occurs at the intersection of the dark solid 
lines in Figure \ref{fig:collide}.  However, it is much simpler 
to treat the collision as if it occurs at the intersection of the
future lightcones of the nucleation events, corresponding to the
intersection of dark dotted lines in Figure \ref{fig:collide}.  
This is an accurate approximate for thin bubble walls and/or 
late bubble collisions.  We focus on the possibility that the 
bubble collision leaves some imprint on the CMB, in which case 
we are interested in the intersection of the past lightcone of 
the observer, the hypersurface of recombination, and the future 
lightcone of the colliding bubble nucleation event.  

Note that we consider only those bubble collisions that create 
small perturbations to the bubble geometry within the observer's 
past lightcone.  One can consider, for instance, the collision 
between two bubbles of the same type as our own, where after the
collision the energy in the bubble walls is converted into 
radiation, which rapidly redshifts away during inflation within 
the bubble.  Meanwhile the collision may affect, for instance, 
the starting value of the inflaton, allowing effects of the 
collision to persist until late times.  One can also imagine 
collisions between a bubble like ours and a bubble with much 
larger vacuum energy, if the post-collision domain wall between 
bubbles rapidly accelerates away from the observer, leaving the 
interior of the observer's bubble largely only weakly perturbed 
by the collision.  For more detailed descriptions see 
e.g.~\cite{worldscollide,AJreview}.   

The intersection of the past lightcone of the observer with the 
hypersurface of recombination corresponds to the surface of last 
scattering.  As we neglect late-time anisotropies, we can describe 
the relevant late-time geometry using the line element
\bea
ds^2=a^2(\eta)\!\left[ -d\eta^2+dr^2+r^2d\theta^2
+r^2\sin^2(\theta)\,d\phi^2\right] ,\,\,\, 
\label{FRW}
\eea
the coordinates of which can be related to the global bubble
coordinate system (in the spacetime region where both charts 
are accurate) according to
\beq
\xi=r\sin(\theta)\,, \quad \phi=\phi\,, 
\quad B\Hd z=r\cos(\theta) \,,
\label{relations}
\eeq
where we have used $b(\eta)/a(\eta)=B\Hd$, as indicated by the 
asymptotic behavior of (\ref{sfa}) and (\ref{sfb}).  Here the
angular coordinate $\theta$ covers the interval $0<\theta<\pi$.

Exploiting some symmetries of the collision geometry, we can use 
boost and $z$-translation invariance to place the observer at 
$(\xi,\,z)=(0,\,0)$.  The surface of last scattering then 
corresponds to the two-sphere with radial coordinate 
$r_\star=\eta_0-\eta_\star$, on the hypersurface $\eta=\eta_\star$, 
where $\eta_0$ is the time of the observer and $\eta_\star$ is 
the time of recombination.  Here we have assumed 
$\eta_0-\eta_\star<\pi B\Hd$, so that the surface of last 
scattering does not wrap entirely around the closed $z$ dimension, 
as indicated by unsuccessful searches in the CMB for a closed 
spatial dimension \cite{O-CTZH}.  Note that the surface of 
last scattering covers only the region 
$0\leq \xi_\star \leq r_\star\ll 1$.

In addition to the above boost, we can use $\sigma$- and 
$\rho$-translation invariance to place the nucleation of the 
observer's bubble at $(\sigma,\,\rho)=(-\Hp^{-1},\,0)$.  Meanwhile
we denote the location of the colliding bubble nucleation event as 
$(\sigma,\,\rho,\,\phi) = (\sigma_{\rm n},\rho_{\rm n},\, 0)$, 
where 
\beq
\Hp(\rho_{\rm n}-\sigma_{\rm n})>1 \,, \qquad
\Hp(\rho_{\rm n}+\sigma_{\rm n})>-1 \,, 
\label{locationconditions}
\eeq
so that the colliding bubble nucleates outside of the observer's
bubble, and vice versa.  The boundary of the future lightcone of
the colliding bubble is then 
\bea
\rho_{\rm c}(\sigma,\,\phi,\,z) = \rho_{\rm n}\cos(\phi)\pm
\sqrt{(\sigma-\sigma_{\rm n})^2-\rho_{\rm n}^2\sin^2(\phi)}\,,\,\,
\label{fl1}
\eea
where it is implicit that the lightcone exists only where 
$\rho_{\rm c}$ is real.  Note that $\rho_{\rm c}$ is independent 
of $z$, due to the bubble collision occurring simultaneously 
across all $z$.  

As we have remarked, the instanton boundary conditions allow us
to extend the coordinates of the parent vacuum into the bubble, 
where at some fiducial time $-\eta_1\gg \ln(\Hp/\Hd)$ we can 
transform the coordinates of the lightcone using (\ref{trans1}) 
and (\ref{trans2}).  The result is a curve in the $(\xi,\,\phi)$ 
plane, 
\bea
\xi(\phi) = \ln\!\left[\frac{1+2\Hp\sigma_{\rm n}
-\Hp^2\!\left(\rho_{\rm n}^2-\sigma_{\rm n}^2\right)}
{1-2\Hp\rho_{\rm n}\cos(\phi)
+\Hp^2\!\left(\rho_{\rm n}^2-\sigma_{\rm n}^2\right)}\right]
-\eta_1 \,,\,\,\,
\label{initialcurv}
\eea
valid for all $-\eta_1\gg \ln(\Hp/\Hp)$, but beyond which the 
future lightcone is complicated to develop.  On the other hand,
the local radius of curvature of (\ref{initialcurv}) is of 
order the radius of curvature of the bubble; meanwhile we are 
only interested in the portion of the future lightcone that 
will intersect the surface of last scattering, the radius of 
which is empirically much smaller than this radius of 
curvature, $0\leq \xi_\star\ll 1$.  We can therefore  
approximate the relevant portion of the lightcone as a line, 
corresponding to the tangent of the future lightcone evaluated 
at $\phi=0$.  In terms of the bubble coordinates this gives
\beq
\xi_{\rm c}(\eta,\,\phi,\,z) = \frac{2\tanh^{-1}\!
\big[\Hp(\rho_{\rm n}+\sigma_{\rm n})\big] -\eta}
{\cos(\phi)} \,.
\label{colliderlightcone}
\eeq    
Again, the future lightcone spans all values of $z$.

The intersection of the surface of last scattering and the 
future lightcone of the colliding bubble nucleation event then 
corresponds to the region  
\beq
\xi_{\rm c}(\eta_\star,\,\phi,\,z) \geq 
\xi_\star(\eta_\star,\,\phi,\,z) \,,
\eeq
where as before $\xi_\star=r_\star\sin(\theta)$ and 
$\eta_\star=\eta_0-r_\star$.  The angular coordinates 
$(\theta,\,\phi)$ on the two-sphere of last scattering 
therefore satisfy  
\beq
\cos(\phi)\sin(\theta) \geq \frac{2}{r_\star}\tanh^{-1}\!
\big[\Hp(\rho_{\rm n}+\sigma_{\rm n})\big] 
-\frac{\eta_0}{r_\star}+1 \,,\,
\label{m&m}
\eeq
which has solutions when 
$\eta_0>2\tanh^{-1}\left[\Hp(\rho_{\rm n}+\sigma_{\rm n})\right]$. 
Note that the angular coordinates satisfying (\ref{m&m}) 
correspond to a disk centered at $(\theta,\,\phi)=(\pi/2,\,0)$.

The disk-shaped profile of the region satisfying (\ref{m&m}) is 
in part a consequence of large-colliding-bubble-lightcone 
approximation of (\ref{colliderlightcone}), and in part a 
consequence of isotropy approximation of (\ref{FRW}).  In both 
cases, however, the corrections to these leading-order results 
are suppressed by the present day curvature parameter 
$\Omega_{\rm c}^0$, which is observationally constrained to be 
very small, and in the context of anisotropic bubble nucleation 
likely to be much smaller; see the discussion surrounding 
(\ref{rstar}).  Furthermore, the precise shape of a given 
collision-affected region on the CMB will be blurred by 
super-imposed inflationary perturbations.  For these reasons
we consider the above approximations to be sufficient.

While we have so far taken the colliding bubble to nucleate at 
$\phi=0$, with more than one such bubble we cannot freely choose 
the $\phi$ coordinate of each nucleation event.  Nevertheless, 
rotational symmetry indicates that the solution for a colliding 
bubble nucleating at $\phi=\phi_{\rm n}$ simply corresponds to 
taking $\phi\to\phi-\phi_{\rm n}$ in (\ref{m&m}).  

\begin{figure}[t!]
\includegraphics[width=0.20\textwidth]{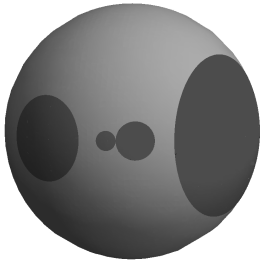} 
\caption{\label{fig:globe}Disks corresponding to the regions 
affected by four bubble collisions, on the two-sphere of an 
observer's CMB sky.}
\end{figure}

Note that, crucially, there is no corresponding generalization of 
the $\theta$-dependence of (\ref{m&m}), since each colliding bubble 
nucleation event occurs simultaneously across all values of $z$.
This implies that the regions affected by such bubble collisions
are always centered at $\theta=\pi/2$, a great circle on the CMB 
sky of the observer.  (The local isotropy of (\ref{FRW}) is broken 
by the choice of preferred coordinate system (\ref{relations}); 
hence the preferred value $\theta=\pi/2$.)  This is illustrated in 
Figure \ref{fig:globe}.  Absent any other observations breaking the 
local background isotropy seen by the observer, two bubble 
collisions are required to determine the preferred $z$ direction 
(the plane with $\theta=\pi/2$), and thus three collisions are 
necessary to verify/falsify the anisotropic bubble nucleation 
hypothesis.

\section{Distribution of bubble collisions}
\label{sec:dist}

The distribution of the number, sizes, and positions of bubble 
collisions in an observer's sky has already been computed in the 
case of (3+1)-dimensional parent and daughter vacua (see 
e.g.~\cite{GGV,AJS,FKNS,Dahlen,AJreview}), and it is 
straightforward to apply the relevant techniques to the present 
scenario.  In particular, our analysis follows very closely 
\cite{FKNS}, and the reader is directed to there and to 
\cite{AJreview} for details that are glossed over in the interest 
of brevity below.  Note that with respect to the distribution of
positions of regions affected by bubble collisions, we have 
already deduced that they are all centered at $\theta=\pi/2$ 
(see Section \ref{sec:collision}).  Therefore we here focus on 
the distribution with respect to the $\phi$ coordinate.  

The calculation involves integrating over all possible nucleation
sites of a colliding bubble, with the parent vacuum idealized to 
cover all of the spacetime outside of the bubbles.  To regulate 
the diverging spacetime volume of the parent vacuum, it is 
customary to draw a spacelike hypersurface at 
$\sigma=\sigma_{\rm in}$, below which it is assumed there are 
no bubble nucleations, and in the end take 
$\sigma_{\rm in}\to -\infty$ \cite{GGV}.  The translations that 
were used to place the observer's bubble nucleation event at    
$(\sigma,\,\rho)=(-\Hp^{-1},\,0)$ merely shift the hypersurface 
$\sigma=\sigma_{\rm in}$.  Furthermore, rotational symmetry allows 
us to define the origin of the $\phi$ coordinate so that the 
observer sits at $\phi=0$ (a colliding bubble then in general 
nucleates at some azimuthal angle $\phi=\phi_{\rm n}$).  On the 
other hand, the boost that was used to translate the observer from 
an arbitrary position $(\xi,\,\phi)=(\xi_0,\,0)$ to $(0,\,0)$ 
corresponds to a non-trivial rotation of the hypersurface 
$\sigma=\sigma_{\rm in}$.  

To understand the effects of this rotation, note that the line 
element of the bubble geometry (\ref{openFRW}) is induced on a 
four-dimensional hypersurface embedded in a (5+1)-dimensional 
Minkowksi space,
\beq 
ds^2=-dt^2+du^2+dv^2+dw^2+dx^2+dy^2 \,,
\eeq 
given the constraint equations
\bea
t \!&=&\! a(\eta)\,\cosh(\xi) \\
u \!&=&\! b(\eta)\,\cos(z) \\
v \!&=&\! b(\eta)\,\sin(z) \\
w \!&=&\! f(\eta) \\
x \!&=&\! a(\eta)\,\sinh(\xi)\,\cos(\phi) \\
y \!&=&\! a(\eta)\,\sinh(\xi)\,\sin(\phi) \,, 
\eea
where the function $f(\eta)$ corresponds to a solution of the 
differential equation
\beq
\dot{f}^2= a^2 + \dot{a}^2 -\dot{b}^2\,.
\eeq 
Then the arbitrary point $(\xi,\,\phi)=(\xi_0,\,0)$ is translated
to the point $(0,\,0)$ by the boost
\bea
t' \!&=&\! \gamma(t-\beta x) \label{boost1}\\
x' \!&=&\! \gamma(x-\beta t) \,, \label{boost2}
\eea  
with the other embedding coordinates unchanged, if we take 
$\gamma=\cosh(\xi_0)$ and $\beta=\tanh(\xi_0)$.  

To understand the effect of this boost on the hypersurface 
$\sigma=\sigma_{\rm in}$, we should embed the coordinates of the 
parent vacuum in the above (5+1)-dimensional Minkowksi space.  
However, the above boost will take the hypersurface 
$\sigma=\sigma_{\rm in}\to-\infty$ to points on the de Sitter 
hyperboloid not covered by the flat slicing, so we first 
transform to a new coordinate system.  A particularly convenient
slicing of the de Sitter submanifold has line element
\bea
ds^2 = A^2(X)\!\left[-dT^2\!+dX^2\!+\cosh^2(T)\,d\phi^2\right]\!
+\!B^2dz^2 ,\,\,\,
\label{newdS}
\eea
where the ``scale factor'' is now $A(X)=\Hp^{-1}{\rm sech}(X)$, 
and the coordinates $T$ and $X$ run from $-\infty$ to $\infty$.
Where the two charts overlap, the above coordinates are related 
to the flat de Sitter ones by
\bea
-\!\Hp\sigma \!&=&\! \frac{\cosh(X)}{\sinh(T)-\sinh(X)} \label{ST1}\\
\Hp\rho \!&=&\! \frac{\cosh(T)}{\sinh(T)-\sinh(X)} \,,
\eea
with trivial identification of $\phi$.  The (5+1)-dimensional 
Minkowksi embedding is determined by the constraints
\bea
t \!&=&\! \Hp^{-1}\sinh(T)\,\,{\rm sech}(X) \\
u \!&=&\! B\,\cos(z) \\
v \!&=&\! B\,\sin(z) \\
w \!&=&\! -\Hp^{-1}\tanh(X) \\
x \!&=&\! \Hp^{-1}\cosh(T)\,\,{\rm sech}(X)\,\cos(\phi) \\
y \!&=&\! \Hp^{-1}\cosh(T)\,\,{\rm sech}(X)\,\sin(\phi) \,. 
\eea

\begin{figure}[t!]
\includegraphics[width=0.25\textwidth]{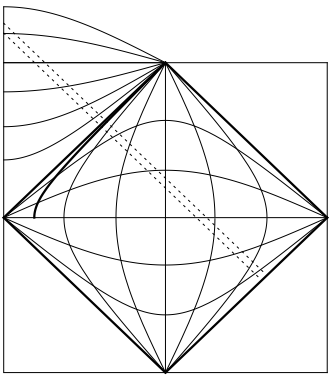} 
\caption{\label{fig:chart}The foliation (\ref{newdS}) of the 
de Sitter submanifold, corresponding to inside the central
diamond-shaped region.  Solid spacelike/timelike curves within
this diamond represent constant $T$/$X$ hypersurfaces, with the
limits $T\to\pm\infty$ corresponding to the top/bottom of the 
diamond, and $X\to\pm\infty$ corresponding to the right/left 
sides of the diamond.  The dashed lines represent hypothetical
curves $\psi=0\,,\pi$.}
\end{figure}

The slicing (\ref{newdS}) of the de Sitter submanifold is 
displayed in Figure \ref{fig:chart}, where it can be seen that 
while the chart does not cover the entire de Sitter hyperboloid, 
it covers all of the spacetime relevant to nucleating a colliding
bubble, except at first glance the bottom-right corner.  However,
by performing the analysis that follows on a de Sitter chart that
covers the entire hyperboloid, one finds that the hypersurface 
$\sigma=\sigma_{\rm in}\to-\infty$ is never boosted into this 
corner, meaning (\ref{newdS}) is sufficient to cover all of the 
spacetime relevant to bubble collisions.  

The hypersurface $\sigma=\sigma_{\rm in}\to-\infty$ translates to 
the hypersurface $\sinh(T)-\sinh(X)=0$ in the slicing 
(\ref{newdS}), or $t+w=0$ in the embedding spacetime.  Performing 
the boost (\ref{boost1})--(\ref{boost2}) gives the bounding 
hypersurface
\bea
\sinh(X_{\rm in})=\gamma\sinh(T_{\rm in})
-\gamma\beta\cosh(T_{\rm in})\cos(\phi_{\rm in}) \,. 
\label{cutoff}
\eea
In these coordinates the observer's bubble can be set to 
nucleate at $(T,\,X)=(0,\,-\infty)$, and the conditions 
(\ref{locationconditions}) merely force the colliding bubble to
nucleate in the spacetime region covered by the chart 
(\ref{newdS}).

The differential number of bubbles nucleating as a function
of the coordinates (\ref{newdS}) can be approximated
\beq
dN = \Gamma dV = \frac{\Gamma}{\Hp^3}\frac{\cosh(T_{\rm n})}
{\cosh^3(X_{\rm n})}\, dT_{\rm n}\,dX_{\rm n}\,d\phi_{\rm n} \,,
\label{dist}
\eeq
where $\Gamma$ is the bubble nucleation rate per unit 
three-volume $V$ in the (2+1)-dimensional effective theory of 
the parent vacuum, and $(T_{\rm n},\,X_{\rm n},\,\phi_{\rm n})$ 
correspond to the coordinates of the nucleating bubble.  
Note that when we use (\ref{dist}) as a measure on the set of
nucleation sites of possibly observable colliding bubbles, we 
make a double-counting error by ignoring the probability that 
a given colliding bubble nucleates within the future lightcone 
of another colliding bubble.  However, given the expected 
smallness of the bubble nucleation rate, $\Gamma/\Hp^3\ll 1$, 
this error is small, for essentially the same reason that 
bubbles do not percolate during eternal inflation 
\cite{GW81,AJS,FKNS}.  Therefore for simplicity we here ignore 
it.

We take interest in the distribution of angular scales $\psi$ 
of the bubble-collision-effected regions in the CMB.  By ``angular
scale $\psi$'' we refer to the disk radius, i.e.~the largest angle 
$\phi$ satisfying (\ref{m&m}) (the largest difference 
$\phi-\phi_{\rm n}$ in the case $\phi_{\rm n}\ne 0$).  In 
terms of $T_{\rm n}$ and $X_{\rm n}$, this is given by (after a 
bit of algebra)
\bea
\cos(\psi) = \frac{1}{r_\star}\big(T_{\rm n}+X_{\rm n}-\eta_0
+r_\star\big) \,.
\label{angularscale}
\eea
Note that, given $X_{\rm n}$ (and $\eta_0$ and $r_\star$), there 
is a one-to-one correspondence between colliding bubble 
nucleation times $T_{\rm n}$ and angular scales in the observer's 
sky $\psi$,
\beq
T_{\rm n}(\psi,\,X_{\rm n}) = 
r_\star\cos(\psi)+\eta_0-r_\star-X_{\rm n} \,.
\label{Tnew}
\eeq 

Since the angular scale $\psi$ is the observable of interest, 
it is convenient to express the differential number of bubble 
nucleations as a function of $\psi$ instead of $T_{\rm n}$.  
Hypersurfaces of constant $\psi$ are parallel to the dashed 
null rays in Figure \ref{fig:chart}, with $0<\psi<\pi$
covering only a sliver of the total spacetime, but all of the
spacetime relevant to observable bubble collisions with effects
spanning less than the observer's entire CMB sky.  Integrating
over $X_{\rm n}$ in the plane $(\psi,\,X_{\rm n})$ involves 
moving down such a null ray until one hits the cutoff, which 
now corresponds to the intersection of (\ref{cutoff}) and 
(\ref{Tnew}), 
\bea
X_{\rm in} = \frac{1}{2}\ln\!\left[
\frac{\varrho(\psi)+\gamma\varrho^2(\psi)
-\gamma\beta\varrho^2(\psi)\cos(\phi_{\rm n})}
{\varrho(\psi)+\gamma +\gamma\beta\cos(\phi_{\rm n})}
\right] ,
\label{Xin2}
\eea
where we have defined
$\varrho(\psi)\equiv \exp\big[r_\star\cos(\psi)+\eta_0-r_\star\big]$
and, as before, $\gamma=\cosh(\xi_0)$ and $\beta=\tanh(\xi_0)$.  
Putting everything together gives
\bea
\frac{dN}{d\psi\, d\phi_{\rm n}} \!\!&=&\!\! \frac{r_\star\Gamma}{\Hp^3}
\sin(\psi)\int_{-\infty}^{X_{\rm in}(\psi,\,\phi_{\rm n})} 
\frac{\cosh\!\big[T_{\rm n}(\psi,X_{\rm n})\big]}
{\cosh^3(X_{\rm n})}\, dX_{\rm n} \phantom{\Bigg[\Bigg]}\!\!\!\nn\\
&=&\!\! \frac{r_\star\Gamma}{\Hp^3} \sin(\psi)
\frac{1\!+\varrho^2(\psi)
\big[1\!+2e^{-2X_{\rm in}(\psi,\,\phi_{\rm n})}\big]}
{\varrho(\psi)\big[1\!+
e^{-2X_{\rm in}(\psi,\,\phi_{\rm n})}\big]^2} \,. 
\phantom{\Bigg[\Bigg]}\!\!\!
\label{main}
\eea

Before proceeding, note that the distribution (\ref{main}) in 
general depends on the angular position of the colliding bubble 
nucleation event, $\phi_{\rm n}$.  This is in addition to the 
anisotropy due to the reduced symmetry of the parent vacuum, 
which expresses itself in the $\theta$-alignment of the regions 
affected by bubble collisions, as described in Section 
\ref{sec:collision}.  Instead the anisotropy with respect to 
$\phi_{\rm n}$ indicates the same ``persistence of memory'' 
effect found by \cite{GGV} in the case of a (3+1)-dimensional 
parent vacuum, which stems from the cutoff hypersurface 
(\ref{cutoff}) breaking the de Sitter symmetries of the parent 
vacuum.  In the case of a (3+1)-dimensional parent vacuum, the 
anisotropy is only significant if $\Hd\approx\Hp$, and even then 
it appears to be unobservable \cite{FKNS}.  Below we find the 
same to hold with our result.

While the distribution function (\ref{main}) may seem 
complicated, under reasonable assumptions it gives way to a 
very simple phenomenology.  Setting aside for the moment the 
constant prefactor $r_\star\Gamma/\Hp^3$, it involves three 
unknowns:  $r_\star$, $\eta_0$, and $\xi_0$.  The observer's 
time $\eta_0$ appears only in addition to 
$r_\star\cos(\psi)-r_\star$, the absolute value of which 
never exceeds $2r_\star$.  While these two terms can in 
principle be comparable, this would require 
$\Hp\lesssim e^{r_\star}\Hd$.  Meanwhile $r_\star$ is 
constrained to be small---see the discussion surrounding 
(\ref{rstar})---so it seems much more plausible 
that $\eta_0\gg r_\star$, in which case 
$\varrho(\psi)\approx e^{\eta_0}\approx \Hp/\Hd$ and 
\beq
\frac{dN}{d\psi}\propto \sin(\psi) \,.
\eeq 
This agrees with the distribution of angular scales in the 
case of a (3+1)-dimensional parent vacuum \cite{AJS,FKNS}.

The observer could in principle sit at any coordinate 
$0\leq \xi_0<\infty$ (before the boost that translates the 
observer to the origin), so one might expect a typical observer 
to reside at $\xi_0\gg 1$.\footnote{By ``typical'' we mean 
randomly selected from a reference class of similar observers 
(with similar environments), including other observers in 
what we have called the observer's bubble, and observers in 
other bubbles of the same type as the observer's bubble but 
nucleating elsewhere in spacetime.  For the purposes of this
paper a ``similar'' observer is any one residing at time 
$\eta_0$ in a bubble consistent with the cosmological 
assumptions above.}  
In fact the location of a typical observer in an 
eternally-inflating multiverse depends on how one regulates 
the diverging spacetime volume.  What spacetime measure is 
appropriate to this or any other prediction is still an open 
question, however the measures receiving the most recent 
attention all predict that typical observers reside at 
$\xi_0\sim {\cal O}(1)$.\footnote{One can discriminate from
among the various measure proposals by searching for 
phenomenological pathologies.  Two pathologies in particular,
Boltzmann brain domination \cite{DKS02,Albrecht,Page1} and 
runaway inflation \cite{FHW,GV05,GS}, seemingly rule out 
measures that predict a uniform distribution of observers 
over the coordinate $\xi$.  The measures known to survive
these pathologies, the causal patch measure \cite{B06}, 
scale-factor cutoff measure \cite{Starobinsky,L06,DSGSV}, and 
comoving probability measure \cite{L06}, all distribute 
observers roughly uniformly according to the flat de Sitter 
comoving congruence of the parent vacuum, which corresponds 
to bubble coordinates $\xi$ predominantly of order unity.}
This by itself does not allow for much simplification of 
(\ref{main}), however if we take one step further and
assume $\Hp/\Hd\gg e^{\,\xi_0}$, the dependence of the 
distribution on $\phi_{\rm n}$ becomes negligible, and 
we obtain
\beq
\frac{dN}{d\psi\, d\phi_{\rm n}} \approx
\frac{r_\star\Gamma}{\Hp^3}\frac{\Hp}{\Hd}\sin(\psi)\,.
\label{main2}
\eeq
Indeed, even in the case where $\xi_0$ is large, careful 
study of (\ref{main}) indicates that the distribution is only 
significantly changed from (\ref{main2}) within a narrow 
region of width $\Delta\phi_{\rm n}\sim\Hd/\Hp$ about the
point $\phi_{\rm n}=0$.  

We can now consider the total number of bubble collisions 
with angular scale smaller than the CMB sky, which is
found by integrating (\ref{main}) over the intervals 
$0<\psi<\pi$ and $-\pi<\phi_{\rm n}<\pi$.  When 
$\Hp/\Hd\gg e^{\,\xi_0}$, the angular dependence is simple, 
and we find
\beq
N \approx 4\pi r_\star\frac{\Gamma}{\Hp^3}\frac{\Hp}{\Hd} \,.
\label{finalN}
\eeq
This result is not significantly changed in the case where
$\xi_0$ is large, so long as $\Hp/\Hd\gg 1$, because the
number of bubble collisions coming from the affected 
part of the distribution is suppressed by a factor 
$\Delta\phi_{\rm n}/2\pi\sim\Hd/\Hp$.  The case 
$\Hp\approx\Hd$ is more complicated, but (\ref{finalN}) 
captures the overall scale of the result.

The factor $\Gamma/\Hp^3$ is the dimensionless decay rate 
of the parent vacuum, and can be crudely approximated as
\beq
\Gamma/\Hp^3\sim e^{-S_{\rm E}} \,,
\label{decayrate}
\eeq
where $S_{\rm E}$ is the Euclidean action of the parent 
vacuum \cite{CDL}.  It is not hard to show that
\beq
S_{\rm E} = \sqrt{\frac{\pi^2B^2}{32\,G^3V}} \,,
\eeq  
where $V$ is the parent vacuum energy (of the 
(3+1)-dimensional Lagrangian, see Section 
\ref{sec:background}).  Thus while $S_{\rm E}$ could 
potentially be very large---and $\Gamma/\Hp^3$ strongly 
exponentially suppressed---there is hope for vacuum 
energies and compactification radii not too far below 
the Planck scale, in which case the suppression would 
not be so severe (and in which case the prefactor 
neglected in (\ref{decayrate}) could also become 
important).  

Meanwhile, $r_\star$ is empirically constrained to be much 
less than unity, and is itself exponentially suppressed by 
the degree to which inflation in the bubble exceeds the 
$\sim 60$ $e$-folds necessary to satisfy the observational 
constraints.  At the same time there are indications that 
long periods of slow-roll inflation may be difficult to 
achieve in string theory, and a toy model of inflation in the 
landscape, combined with anthropic selection, gives 
reasonable hope that inflation may not have lasted too long 
within our bubble \cite{FKRMS,BAP,BL09,omega}.  

Countering these two small factors is the ratio $\Hp/\Hd$, 
which could in principle be very large, for instance 
$\sim 10^{30}$ in the case of TeV-scale inflation with a near 
Planck-scale parent vacuum.  Notice however that without this 
factor the number of bubble collisions is necessarily 
typically much less than unity.  Thus it seems the effects
of such bubble collisions are likely to be observed only if 
the parent vacuum energy is very large, the scale of 
inflation in our bubble is relatively small, and inflation in
our bubble does not last too long beyond what is necessary to
make the present geometry approximately flat.  

Finally, recall that the number of bubble collisions coming 
from the region $\Delta\phi_{\rm n}$ affected by anisotropy 
with respect to $\phi_{\rm n}$ is suppressed by a factor 
$\sim\Hd/\Hp$.  We can now see that this factor cancels the 
necessary enhancement to $N$ mentioned above, which is why 
the ``persistence of memory'' anisotropy is considered 
unobservable.

\section{Conclusions}
\label{sec:conclusions}

We have studied the properties of bubble collisions in an
eternally-inflating parent vacuum where one of our three large 
spatial dimensions is compact.  In particular, we have focused 
on bubbles formed via tunneling of a metastable modulus, 
corresponding to decompactifying a spatial dimension of the 
parent vacuum, producing bubbles with internal geometries that, 
after another period of inflation, resemble our own FRW 
cosmology.  Such collisions are not unexpected in a multiverse 
populated by the string landscape, though our universe may or 
may not reside in a bubble of this type.

We find that such collisions may leave a distinct signature 
on the CMB sky, as their effects are limited to disks centered
on a single great circle on the two-sphere of last scattering.  
This is very different than the case of a (3+1)-dimensional 
parent vacuum, where the corresponding disks can appear 
anywhere on the CMB sky.  The distribution of angular scales of 
such regions, and of their azimuthal positions, qualitatively 
resemble those in the case of a (3+1)-dimensional parent vacuum.  
In particular, the distribution of angular scales $\psi$ 
follows $dN/d\psi\propto \sin(\psi)$.  The probability that a 
bubble collision both resides in our past lightcone and has 
effects that do not cover the entire CMB sky depends on 
properties of the parent vacuum, the tunneling instanton, and 
inflation in the bubble.  It could be much less than unity but 
there is hope for parameter values that make observing the 
effects of bubble collisions more likely.  

There are many remaining interesting questions.  Perhaps the 
most pressing issue, pertaining to signatures of bubble 
collisions in this and other scenarios, is to determine what 
such collisions would actually look like in terms of CMB 
observables (for instance, what is the temperature profile 
of an affected region).  Also, in the context of anisotropic 
bubble nucleation, we have explored only one type of bubble 
collision---that between two bubbles with one more large 
spatial dimension than an effectively (2+1)-dimensional parent 
vacuum---but one can also consider collisions between bubbles 
like ours 
and lower- and higher-dimensional bubbles, or different 
parent-vacuum geometries altogether.

\acknowledgments

The author thanks Jose Blanco-Pillado for several informative
discussions, Ben Freivogel for helpful discussion, Matthew 
Johnson for an invaluable critique of an earlier manuscript and 
other useful comments, and Matthew Kleban for identifying a 
significant error in an earlier manuscript.  This work was 
supported by the U.S. National Science Foundation under grant 
0855447.

\end{document}